\documentclass[twocolumn,floats,aps]{revtex4}
\usepackage{times,amsmath,graphics,psfrag,latexsym}

\begin{document}
\title{Implications of $T_c$ variation in UBe$_{13}$ for a possible
Fulde-Ferrell-Larkin-Ovchinnikov phase}
\author{H.A. Radovan and R.J. Zieve}
\address{Physics Department, University of California at Davis, Davis, CA 95616}
\author{J.S. Kim and G.R. Stewart}
\affiliation{Physics Department, University of Florida, Gainesville, FL  32661}
\begin{abstract}
We measure the heat capacity of UBe$_{13}$ with an unusually low $T_c$  for a
polycrystal.  We find an upturn in the upper critical field $H_{c2}(T)$ below
about $T_c/2$, much as for higher-$T_c$ samples.  Comparing the critical fields
in our sample and in samples with higher $T_c$'s shows that the low-temperature
limit of $H_{c2}$ is proportional to $T_c(H=0)$, as expected if the upturn
comes from an FFLO phase and strong coupling. 
\end{abstract}
\maketitle

Although discovered in 1983, the heavy-fermion superconductor UBe$_{13}$
\cite{Ott83} still has many incompletely understood properties.  These
include the splitting of the superconducting transition upon doping the
U sites with thorium, and the power law temperature dependences of various
quantities in the superconducting phase.
Another issue is the
temperature dependence of the upper critical field $H_{c2}(T)$.  Unique
among heavy fermion superconductors, UBe$_{13}$ shows a clear upturn in
$H_{c2}(T)$ at about half the superconducting transition temperature, with
$H_{c2}(T=0)$ far exceeding the Clogston paramagnetic limit \cite{Glemot99}.
Indeed, the only other material with similar $H_{c2}$ behavior is the
recently discovered UGe$_2$, which superconducts under pressure
\cite{Sheikin01}. 

The two most detailed explanations of the upturn involve a mixture of two
representations in the order parameter \cite{Fomin} and an enhancement of
the paramagnetic limit of $H_{c2}(T=0)$ by strong coupling \cite{Thomas96}.
Here we show that the behavior of $H_{c2}(T)$ in samples of different $T_c$
is consistent with the latter explanation.  
Thomas et al. \cite{Thomas96} fit $H_{c2}(T)$  with three adjustable parameters:
the slope $dH_{c2}/dT(T_c)$, which probes the orbital field limit; the
gyromagnetic ratio $g$ of the quasiparticles; and the strong coupling parameter
$\lambda$.  Matching the upturn requires the further assumption of a
Fulde-Ferrell-Larkin-Ovchinnikov (FFLO) state \cite{Fulde64, Larkin65}. 

Several limits influence
the low-temperature critical field.  The
orbital limit, $H_{c2}^{orb}(T)=H_{c2}^{orb}(0)(1-(\frac {T}{T_c})^2)$,
determines the critical field in most superconductors.  The
huge effective masses in heavy fermion superconductors make spin effects far
more important than usual, although the orbital influence continues to dominate
very close to $T_c$.  A second field is the paramagnetic limit $H_{c2}^p$, also
known as the Clogston or Pauli limit. For spin singlet pairing, the
superconducting condensation energy competes with the magnetic energy tending to
align the quasiparticle spins.  In a single-particle excitation model, this
leads to an abrupt quenching of superconductivity when the applied field reaches
the Clogston limit $H_{c2}^p(T)=\frac{\sqrt{2}\Delta(T)}{g\mu_B}$, where
$\Delta$ is the superconducting energy gap and $\mu_B$ is the Bohr magneton. 
Yet the Clogston limit is not the final word on spin limitations of $H_{c2}$. 
At high magnetic fields and $T$ less than about $T_c/2$, a
Fulde-Ferrell-Larkin-Ovchinnikov (FFLO) state \cite{Fulde64, Larkin65} may
appear in clean superconductors.  $H_{c2}^p$ is derived in a single excitation
picture, but multiple excitations interact by reducing the energy gap.  In the
FFLO state pairs are broken over an entire portion of the Fermi surface, with
the resulting quasiparticles aligned by spin.  The pairing over only part of the
Fermi surface leads to an anisotropic energy gap with planar nodes perpendicular
to the field.  The energy difference between the spin-up and spin-down branches
of the Fermi surface gives a finite center-of-mass pairing $Q$.  Further energy
considerations show that in fact the order parameter amplitude undergoes a
real-space modulation.  Since the FFLO state has partial spin alignment, its
critical field $H_{c2}^{FFLO}$ exceeds $H_{c2}^p$.

\begin{figure}[b]
\begin{center}
\psfrag{C/T (J/mol K^2)}{\scalebox{2.4}{$C/T$ (J/mol K$^2$)}}
\scalebox{0.41}{\includegraphics{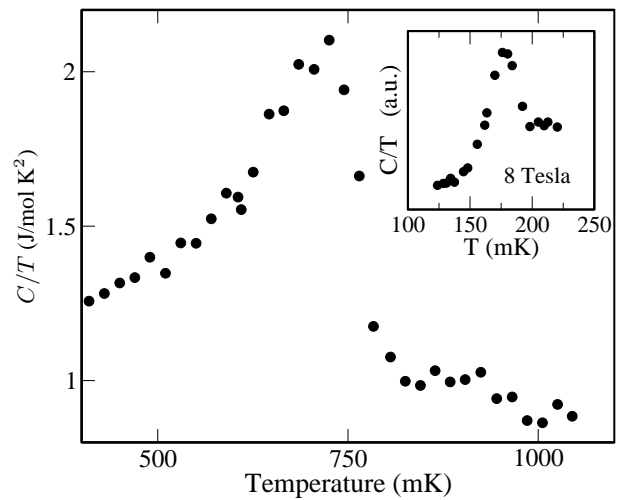}}
\caption{\small Heat capacity of our UBe$_{13}$ sample.
The main figure is with no applied field, the inset
at 8 Tesla.}
\label{f:8tesla}
\end{center}
\end{figure}

These low-temperature critical field behaviors have different dependence on the
sample's transition temperature.  The Clogston limit is  proportional to the energy
gap, leading to a proportionality to the zero-field transition temperature as well. 
The strong coupling parameter enters here through its effects on the gap. Similarly,
the energetics of the FFLO state scale with the energy gap, and hence with $T_c$, so 
$H_{c2}^{FFLO}(t=T/T_c)$ should also be proportional to $T_c$.  On the other hand,
$H_{c2}^{orb}$ depends on the coherence length $\xi$, and its scaling with $T_c$ is not
straightforward.

Fortunately, another unusual feature of UBe$_{13}$ allows us to test this
scaling. Reported values for the onset of the superconducting transition vary by
more than 200 mK, with high-quality crystals (as judged by the transition width)
at both extremes of the range. Langhammer et al. suggest that two types of
UBe$_{13}$ exist: L(low)-type, with $T_c$ around 750 mK; and H(high)-type, with
$T_c$ from 850 to 950 mK \cite{Langhammer98}.  Furthermore, polycrystals are
supposedly all H-type, while single crystals fall into both categories.  
Here we show the $H_{c2}$ upturn in a polycrystal with a particularly low
$T_c$.  The only sample with such a low transition temperature previously
studied had not shown the $H_{c2}$ upturn \cite{Langhammer98}.
Comparing the shape of the upturn in samples of different $T_c$ supports
the FFLO interpretation of the upturn.

Our sample was prepared by arc melting in an argon atmosphere and subsequent
annealing at 1400$^\circ$ C for 1000 hours in a beryllium atmosphere
\cite{Kim91}.  It is polycrystalline with a typical grain size of 100
$\mu\mbox{m}$, as seen in a Philips CM-30 transmission electron microscope.  We
use a relaxation method to measure heat capacity $C(T)$, shown in Figure
\ref{f:8tesla}, from 100 mK to 900 mK. The superconducting transition width is
small, about 50 mK, speaking to a high sample quality despite the low $T_c$.

The inset of Figure \ref{f:8tesla} shows heat capacity in our highest
magnetic field, 8 Tesla, measured on a small piece cut from the same sample
as that for the main figure.  This piece is less than 20 microns thick and
less than 1 mm in lateral dimensions.  Because of its small size, the
background heat capacity is significant but the transition remains clearly
visible.  The piece was made thin for irradiation with high-energy uranium
ions to study the influence of lattice disorder on the superconducting
phases \cite{Radovan01}.  Although the irradiation shows no measurable
effects up to a density of $10^{15}$ tracks per square meter, we performed
various measurements as a function of applied field while searching for
matching effects at fields where the defect density equals the vortex
density. The results presented here are from the irradiated sample which we
studied in the most detail, with track density $5\times 10^{13}$/m$^2$. 
This sample's zero-field $T_c$ and transition width are identical to those of
the unirradiated samples.

\begin{figure}[thb]
\begin{center}
\psfrag{Hc2 (T)}{\scalebox{2.4}{$H_{c2}$ (Tesla)}}
\scalebox{0.45}{\includegraphics{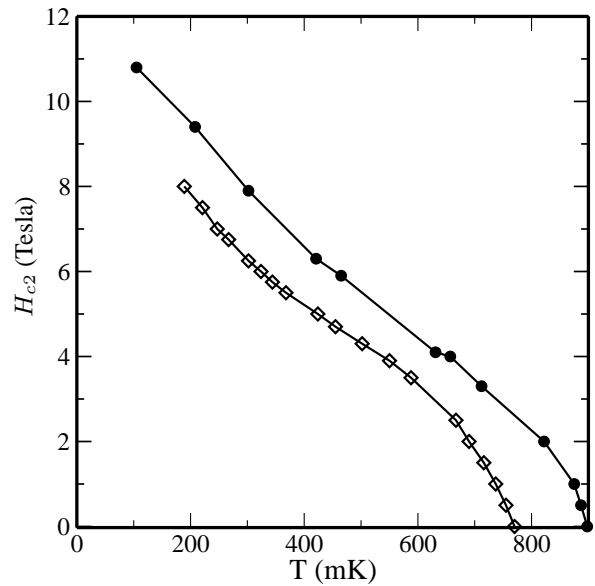}}
\caption{\small Upper critical field as a function of temperature for our sample
(diamonds) and from Reference \cite{Langhammer98} (circles).}
\label{f:Hc2}
\end{center}
\end{figure}

Figure \ref{f:Hc2} combines our own data with previous measurements on a
higher-$T_c$ sample \cite{Langhammer98}.  Since the criterion for identifing
$T_c$ can change the curve shape of $H_{c2}(T)$ \cite{Thomas96}, both our data 
and the high-$T_c$ comparison curve use $T_c$ as the transition midpoint of
heat capacity measurements.  The transition widths of the two samples are also
comparable. Both curves display the characteristic upturn in $H_{c2}$.  The
temperature of the upturn tracks the zero-field $T_c$, remaining near $T_c/2$.
The shift we observe in the temperature of the upturn with $T_c$ is consistent
with an FFLO explanation.

Near $T_c$, resistance measurements at fields below 0.28 kG give
$\frac{dH_{c2}}{dT}(T_c)=6.5$ Tesla/K for our sample, much lower than the 42
Tesla/K found in UBe$_{13}$ samples with $T_c$ near 900 mK \cite{Maple85}.  If
orbital effects dominate near $T_c$, this slope is related to the
zero-temperature orbital limit.  The smaller value in the lower-$T_c$ sample
could indicate a smaller superconducting coherence length.  
It is not clear why $\frac{dH_{c2}}{dT}(T_c)$ depends so strongly on the
sample's $T_c$, but we note that
$\frac{dT_c}{dP}$ has a similar variation, with pressure affecting $T_c$ far
more strongly for a higher-$T_c$ sample
\cite{Miclea02}.

\begin{figure}[tb]
\begin{center}
\psfrag{H (Tesla)}{\scalebox{2.4}{$H$ (Tesla)}}
\psfrag{t=T/Tc}{\scalebox{2.7}{$t=T/T_c$}}
\scalebox{0.45}{\includegraphics{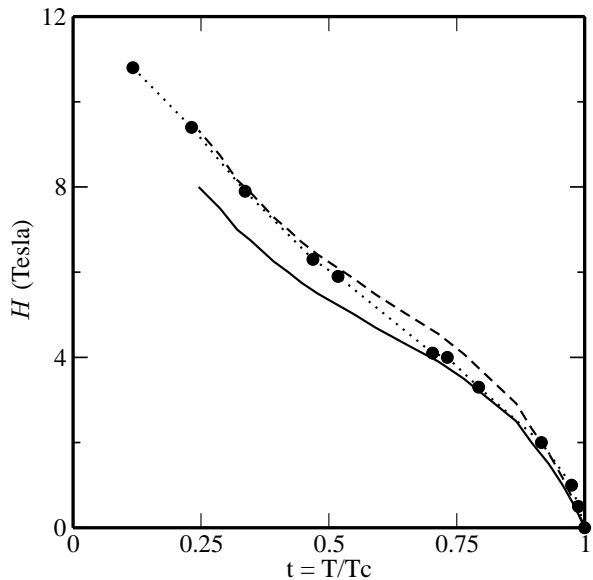}}
\caption{\small Circles and dotted line: upper critical field from heat capacity on
high-$T_c$ sample.  Solid line: our measurements.  Dashed line: same as solid line,
but with field scaled by the same factor as temperature.}
\label{f:Hcscaled}
\end{center}
\end{figure}

At low temperatures, the $H_{c2}$ curves have a natural scaling, as
shown in Figure \ref{f:Hcscaled}.  The circles and dotted line show the
higher-$T_c$ curve from Figure \ref{f:Hc2}, as a function of $t=T/T_c$.  The dashed
line shows our measurements, again as a function of $t$, and with the field
values also scaled by the ratio of the two transition temperatures,
$h=H_{c2}*T_c(\mbox{high})/T_c(\mbox{ours})$. Since these two curves overlay
for $t<0.45$, the low-temperature critical fields for the samples scale with
$T_c$.  If this region is dominated by paramagnetic effects, including
the FFLO enhancement, this scaling is expected. We find a different and
unexplained scaling at intermediate temperatures. The solid curve of Figure
\ref{f:Hcscaled} shows the same data as the dashed line, but with the scaling
applied only to the temperature axis. From $t=0.7$ to near 1, this curve
coincides with the low-$T_c$ data.

With respect to the $H_{c2}$ upturn, Th-doped compounds behave quite
differently from pure UBe$_{13}$. In the U$_{0.97}$Th$_{0.03}$Be$_{13}$, no
upturn occurs in $H_{c2}$ down to $t=0.26$ \cite{Jin96}.  This is consistent
with the upturn indicating an FFLO phase, which is destroyed by disorder in
thorium-doped material. Another possibility is that the upturn in pure
UBe$_{13}$ is a precursor for the second phase transition
\cite{Rauchschwalbe87}. For example, the two upper critical fields in
U$_{0.978}$Th$_{0.022}$Be$_{13}$ merge below about $t=0.59$ \cite{Jin96}. 
Since $H_{c2}(T)$ has different slopes in the two phases, the boundary
between normal and superconducting phases in the $H-T$ plane has an upturn
much like that in pure UBe$_{13}$.  

Because the wide range of $T_c$'s for samples with narrow transitions in
UBe$_{13}$ is so unusual, with its origin not well understood, we make a few
further comments on our sample.  Suggested causes \cite{Langhammer98} for the
$T_c$ variation include aluminum impurities, and variations in the beryllium
content, which must be carefully monitored during the preparation process
\cite{Kim91}.  Previously only single crystals had been reported with such low
transition temperatures, and explanations focused on differences between single
crystal and polycrystal preparation. However, this seems to be a consequence of
reporting $T_c$ as the onset rather than midpoint of the transition. In Table
\ref{t:transitions} we list both midpoints and onsets for our sample and for
those from several previously published works. Except as otherwise noted, we
extracted these numbers from heat capacity graphs in those papers.  Resistive
measurements generally yield slightly higher values, but the large span of
$T_c$'s remains.  We note that the transition {\em midpoints}, for both single
crystals and polycrystals, are widely distributed between 725 mK and 930 mK. 
Since previous polycrystalline samples with low $T_c$'s also had wide
transitions, the $T_c$ {\em onsets} were all near 900 mK.  In our sample, with
its narrow transition, $T_c$ is clearly much lower.  Since the samples with the
narrowest transitions do not have the highest onset temperatures, identifying
the midpoint temperatures is reasonable. 

\begin{table}
\caption{Transition temperatures of UBe$_{13}$ measured by heat capacity
for various samples.  Both the midpoint and onset of the transition are given.}
\label{t:transitions}
\begin{tabular}{cccc}
Midpoint (mK) & Onset (mK) & Width (mK) & Reference\\
Single crystals & & \\
725 & 770 & 90 & \cite{Reinders94}\\
745 & 780 & 70 & \cite{Ramirez99}\\
744 & 768 & 48 & \cite{Langhammer98}\\
790 & 880 & 180 & \cite{Ott83}\\
860 & --  & $<96$ & \cite{Jin94}\\
900 & 920 & 40 & \cite{Langhammer98,Thomas96}\\
907 & 950 & 86 & \cite{Fisk88}\\
Polycrystals & & \\
765 & 790 & 50 & This work\\
830 & 910 & 160 & \cite{Heffner91}\\
830 & 885 & 110 & \cite{Beyermann95}\\
885 & 945 & 120 & \cite{Rauchschwalbe87}\\
915 & 980 & 130 & \cite{Ramirez99}\\
919 & -- & $<40$ & \cite{Jin94}\\ 
930 & 970 & 80 & \cite{Ott86}\\
\end{tabular}
\end{table}

We find an upturn in $H_{c2}(T)$ below about $T_c/2$, a feature previously seen
only in UBe$_{13}$ samples with $T_c$ around 900 mK. Comparing the shape of
$H_{c2}(T)$ for samples with different $T_c$'s meshes well with interpreting
the upturn as an FFLO state in a strong coupled superconductor.  As $T_c$
changes the paramagnetic limit for critical field scales with $T_c$, which we
find for $H_{c2}$ at low temperatures. The orbital limit depends on the
superconducting coherence length and changes with $T_c$ in a less
straightforward way, in agreement with our measurements near $T_c$. Observing
the critical field upturn in a low $T_c$ sample does raise a question about the
role of disorder in causing the wide $T_c$ range. If the upturn indicates an
FFLO phase, which requires a clean limit, then the upturn should be suppressed
in low $T_c$ samples, as it is in thorium-doped UBe$_{13}$.  The resolution
probably depends upon a better understanding of the $T_c$ variation itself. Our
measurements on a polycrystalline UBe$_{13}$ sample with $T_c$ = 770 mK and our
compilation of past work present evidence for the existence of only one type of
UBe$_{13}$ with a range in $T_c$ between 725 mK and 930 mK.  

We thank R. Field for TEM measurements.
This work was supported by NSF under DMR-9733898 (UCD) and by DOE under
DE-FG05-86ER45268 (Florida).

\end{document}